\documentclass{article}
\usepackage{icrctc07,graphicx,amssymb}
\title{On the status of the dip in UHECR spectrum}
\shorttitle{On the status of the dip in UHECR spectrum}
\authors{V.~Berezinsky$^{1}$, A.~Gazizov$^{2}$, S.~Grigorieva$^{3}$ }
\shortauthors{V.~Berezinsky et al.}
\afiliations{$^1$INFN, Laboratori Nazionali del Gran Sasso,
I-67010 Assergi (AQ), Italy \\ %
$^2$B.~I.~Stepanov Institute of Physics of NASB, Independence Avenue 68,
220072 Minsk, Belarus \\ %
$^3$Institute for Nuclear Research of RAS, 60th October Revolution
prosp.\ 7A, 117312 Moscow, Russia } %
\email{venya.berezinsky@lngs.infn.it}

\abstract{The status of the pair-production dip as a spectral
feature, produced by interaction of Ultra High Energy
extragalactic protons with CMB is discussed.}

\begin{document}
\maketitle

Greisen-Zatsepin-Kuzmin (GZK) cutoff \cite{GZK} is the most
spectacular prediction for Ultra High Energy Cosmic Ray (UHECR)
spectrum, which status is still uncertain in the present
observations. As physics is concerned, detection of the GZK cutoff
means discovery of UHE proton interaction with CMB radiation.
Another prediction for interaction of UHE protons with CMB is {\em
pair-production dip}, the spectral feature originated from
electron-positron pair production by extragalactic UHE protons
propagating through CMB: $p+\gamma_{\rm CMB} \rightarrow p+e^++e^-$.
Originally proposed for diffuse spectrum in early work \cite{BG88},
this feature has been studied recently in Refs.\ \cite{BGG,BGGPL}.
An alternative explanation of the observed pair-production dip,
widely discussed now \cite{ankle}, was first put forward in works
\cite{HS85} and \cite{YT} in terms of a two-component model as the
transition from galactic to extragalactic cosmic rays. Being a quite
faint feature, the $e^+e^-$-production dip is not seen well in the
naturally presented spectrum $\log J(E)$ vs.\ $\log E$. The dip is
more pronounced when analyzed in terms of the {\em modification
factor} \cite{BG88,Stanev00}, $\eta(E)=J_p(E)/ J_p^{\rm unm}(E)$,
where $J_p(E)$ is the spectrum calculated with all energy losses
included, and $J_p^{\rm unm}(E)$ is the unmodified spectrum
calculated with adiabatic energy losses only. The observed
modification factor is given by $\eta_{\rm obs} \propto J_{\rm
obs}(E)/E^{-\gamma_g}$, where $J_{\rm obs}(E)$ is the observed
spectrum and $\gamma_g$ is the exponent of the generation spectrum
$Q_{\rm gen}(E_g) \propto E_g^{-\gamma_g}$ in terms of initial
proton energies $E_g$.

\begin{figure*}[ht]
\begin{center}
   \begin{minipage}[ht]{54 mm}
     \centering
     \includegraphics[width=53 mm]{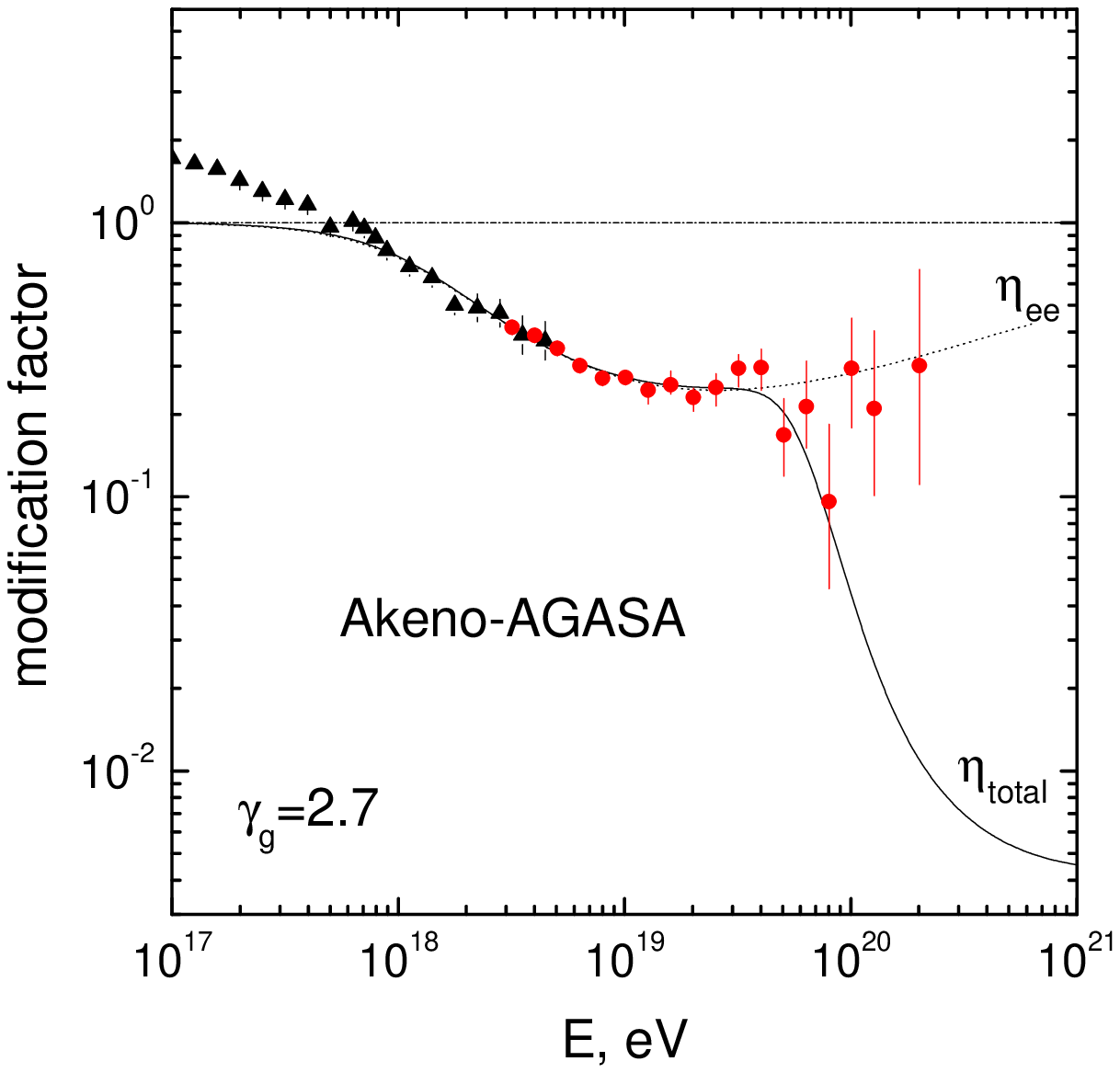}
   \end{minipage}
   \hspace{1mm}
   \vspace{-1mm}
 \begin{minipage}[h]{54 mm}
    \centering
    \includegraphics[width=53 mm]{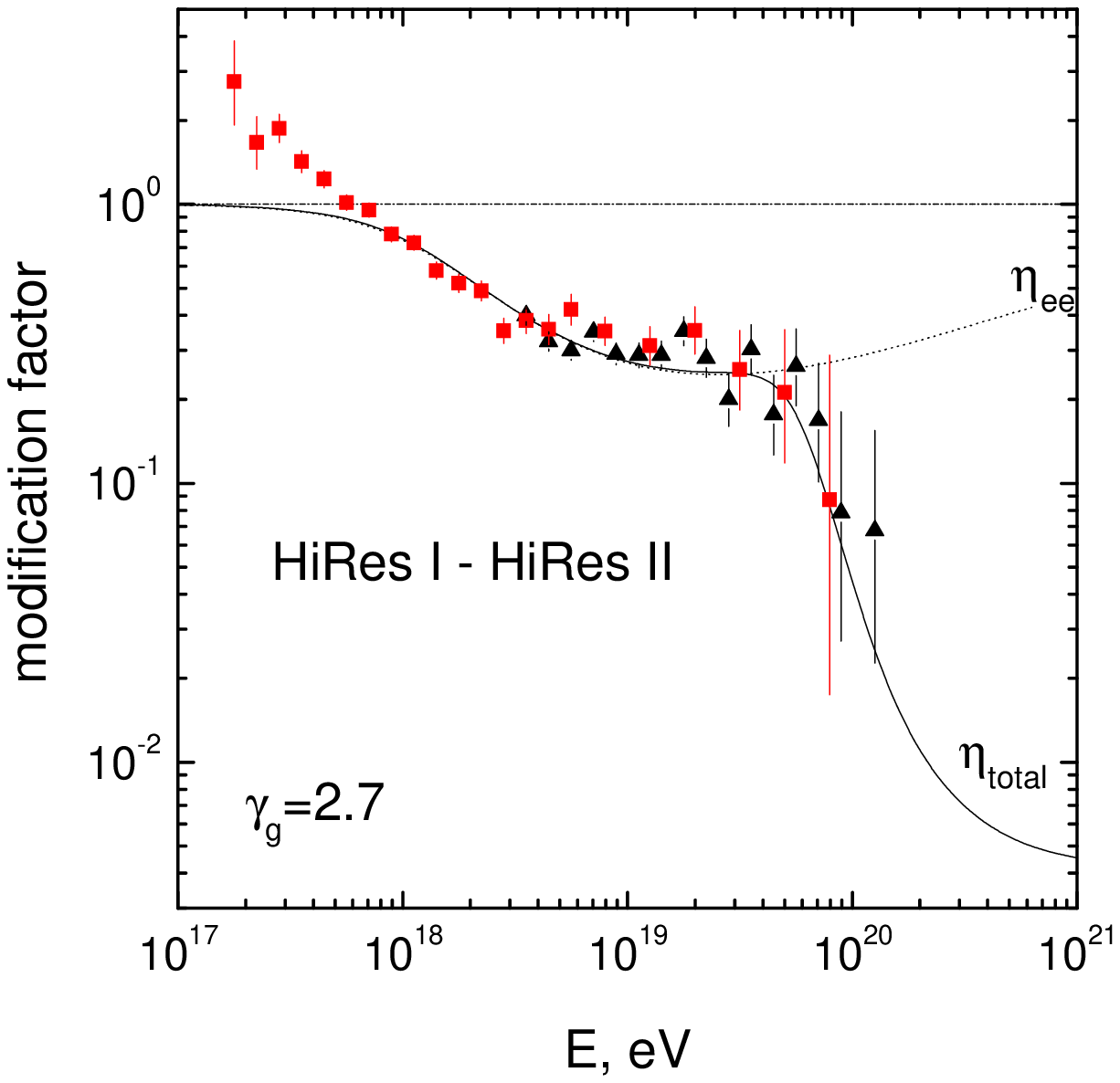}
 \end{minipage}
\medskip
   \begin{minipage}[ht]{54 mm}
     \centering
     \includegraphics[width=53 mm]{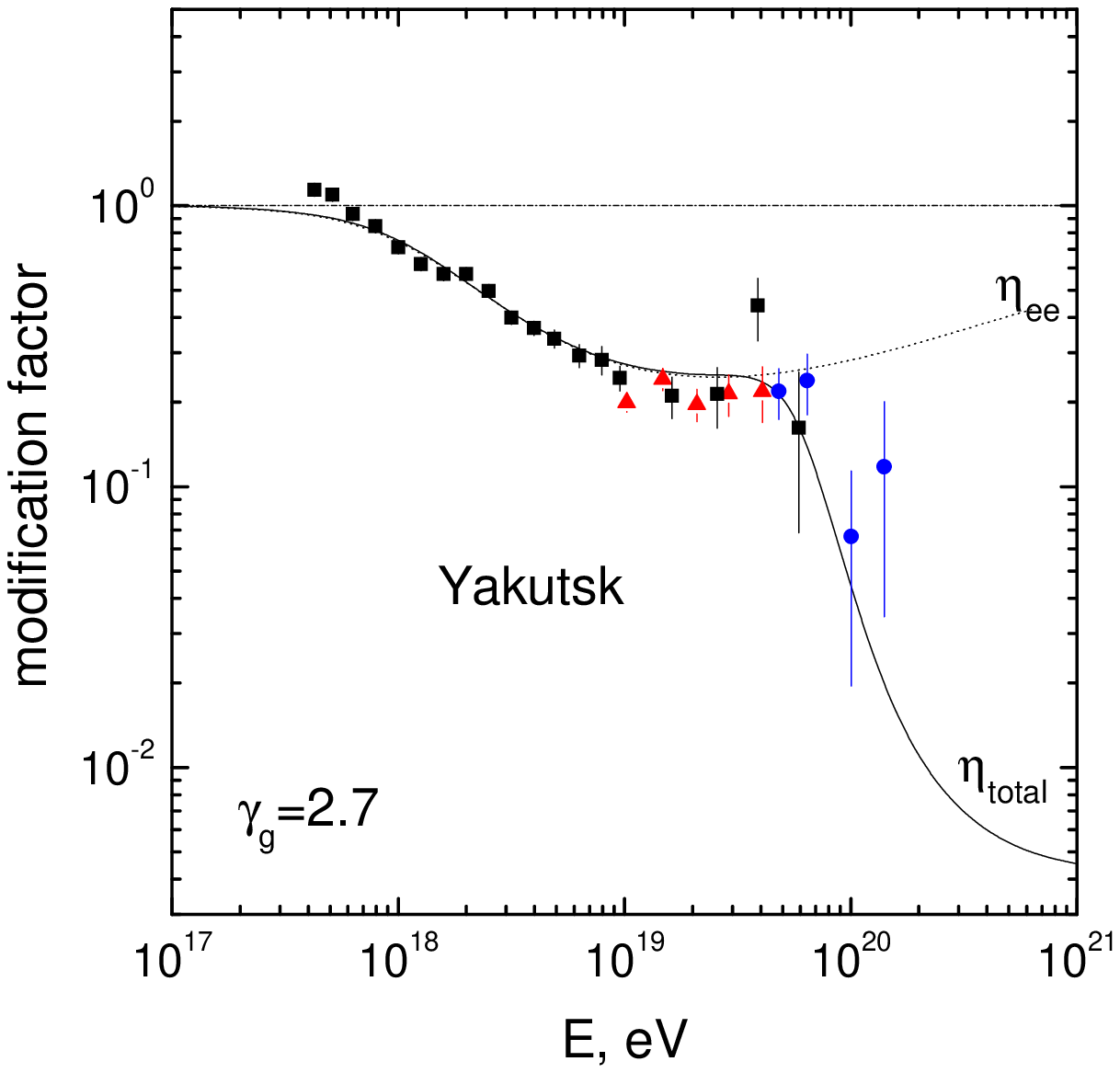}
   \end{minipage}
   \hspace{1mm}
   \vspace{-1mm}
 \begin{minipage}[h]{54 mm}
    \centering
    \includegraphics[width=53 mm]{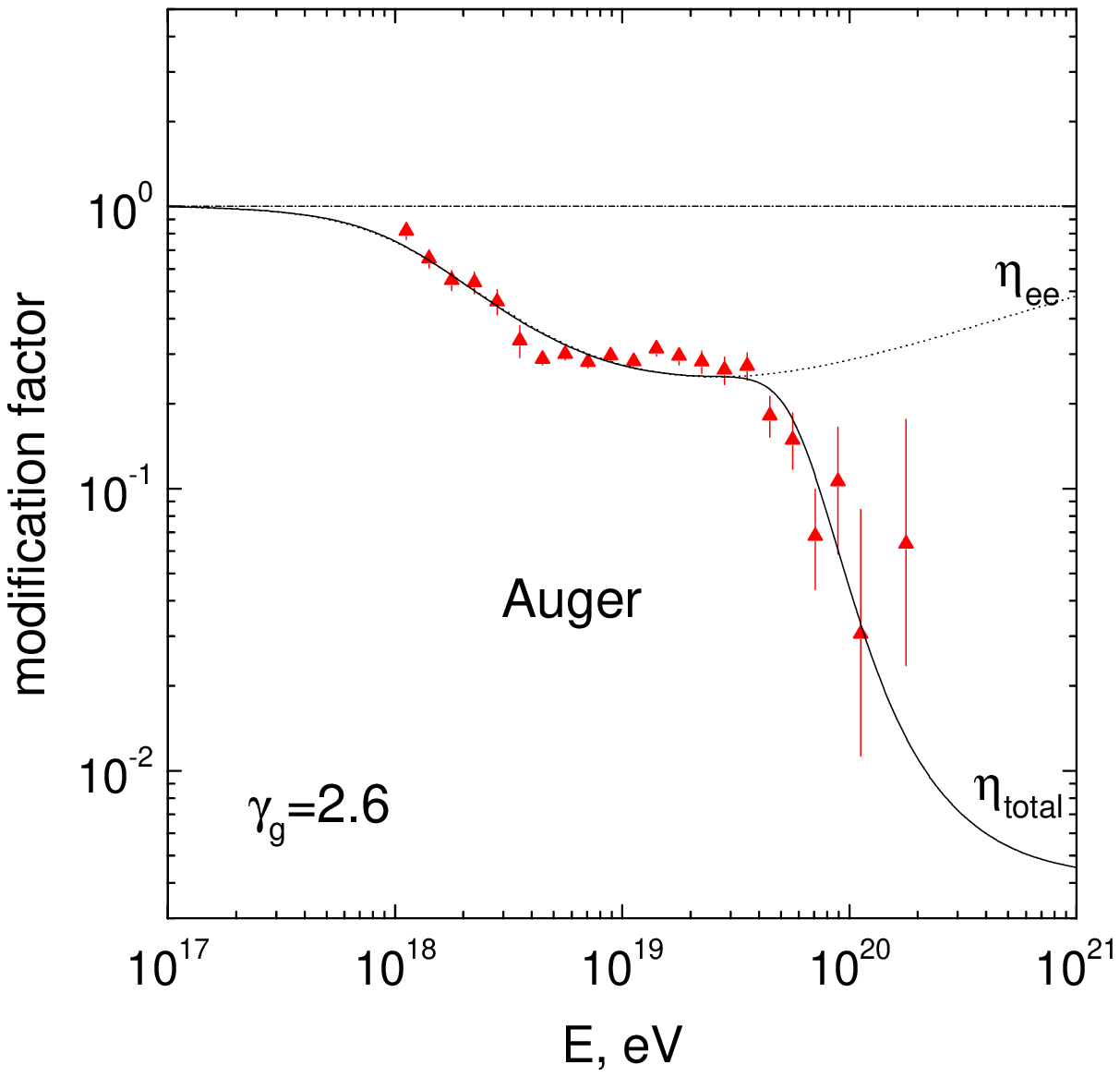}
 \end{minipage}
\end{center}
\vspace{-4 mm}%
\caption{ The predicted pair-production dip in comparison with
  Akeno-AGASA, HiRes, Yakutsk and Auger data \cite{data}.
  The first three experiments confirm dip with good $\chi^2/{\rm
  d.o.f.} \approx 1.0 -1.2$, while the  Auger data are characterized
  by larger $\chi^2/{\rm d.o.f.}$, mainly due to the bin at energy
  $E \approx 45$~EeV.
  The data of Fly's Eye \cite{data} (not presented here) confirm
  the dip as AGASA, HiRes and Yakutsk detectors do.} %
  \label{fig:dips}
\end{figure*}

The pair-production dip is clearly seen in the energy-dependence of
$\eta(E)$ and is reliably confirmed ~\cite{BGG,BGGPL,Aletal} by
observational data, as Fig.~\ref{fig:dips} shows. The comparison of
the predicted dip with observational data includes only two free
parameters: exponent of the power-law generation spectrum $\gamma_g$
(the best fit corresponds to $\gamma_g=2.6 - 2.7$) and normalization
constant to fit the $e^+e^-$-production dip to the measured flux.
The number of energy bins in the different experiments is 20 - 22.
The fit is characterized by $\chi^2/{\rm d.o.f.} = 1.0 - 1.2$ for
AGASA, HiRes and Yakutsk data. For the Auger data $\chi^2/{\rm
d.o.f.}$ is larger mainly due to the low flux in the first energy
bin at $E \approx 45$~EeV where measurements are made with the help
of surface detectors (see Fig.~\ref{fig:dips}).

The theoretical pair-production dip has two flattenings: one at
energy $E_b \approx 1\times 10^{18}$~eV and the other at $E_a
\approx 1\times 10^{19}$~eV. One can see that at $E < E_b$ the
experimental modification factor, as measured by Akeno and HiRes,
exceeds the theoretical modification factor. Since by definition
modification factor must be less than one, this excess signals the
appearance of a new component of cosmic rays at $E < E_b = 1 \times
10^{18}$~eV, and thus the transition from extragalactic to galactic
cosmic rays, starting at energy $E_b$.\\*[1mm]
The second flattening automatically explains\\
the {\em ankle}, the
feature seen in all experiments starting from Haverah Park in the end
of 70s.

The position and shape of the dip is robustly fixed by interaction
with CMB and can be used for energy calibration of the detectors.

\begin{figure*}[ht]
\begin{center}
\includegraphics[width=0.8\textwidth]{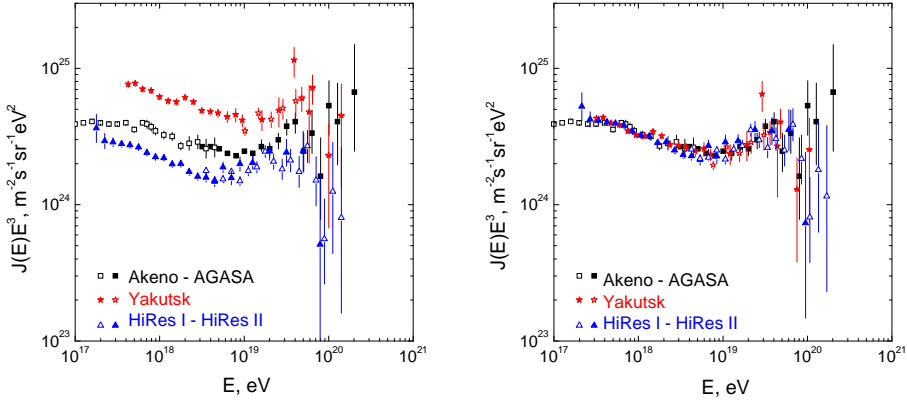}
\end{center}
\caption{The fluxes from  Akeno-AGASA, HiRes and Yakutsk detectors
  before and after calibration by the
  $e^+e^-$-production dip.} %
\label{fig:AgHiYa}
\end{figure*}

The systematic errors in energy measurements are high, from 15\%
in AGASA to 22\% in Auger. To calibrate each detector we shift the
energies by factor $\lambda$ to reach minimum $\chi^2$ in
comparison with theoretical dip. We obtain these factors as
$\lambda_A=0.9$,~ $\lambda_{Ya}=0.75$~ and $\lambda_{Hi}=1.2$~ for
AGASA, Yakutsk and HiRes detectors, respectively. Recently, AGASA
collaboration has reduced their energies by 10\% indeed, based on
reconsideration of  energy determination. After energy
calibration the fluxes given by AGASA, HiRes and Yakutsk detectors
agree with each other in most precise way (see
Fig.~\ref{fig:AgHiYa}). The Auger flux is noticeably below the
flux shown in Fig.~\ref{fig:AgHiYa}.

Concerning the calibration two remarks are in order.

{\em i)} After calibration the discrepancy between AGASA and HiRes
data at the highest energies diminishes to the level of $2.5\;
\sigma$, but the AGASA excess over the theoretical flux with the
GZK cutoff remains statistically significant. The better agreement
between highest energy AGASA and HiRes data implies some trial
theoretical spectrum between AGASA and HiRes data.

{\em ii)} One can see that calibration with help of the
pair-production dip implies decreasing energies measured by
on-ground methods ($\lambda_A=0.9$ and $\lambda_{Ya}=0.75$) and
increasing the energies measured by fluorescent method
($\lambda_{Hi}=1.2$). It might be considered as an indication to
the difference in measuring energies by these two methods.
The predicted shape of the $e^+e^-$-production dip is quite robust
\cite{BGG,Aletal}: it is modified very weakly when the
discreteness in the source distribution and their inhomogeneities
are taken into account, and different regimes of propagation (from
rectilinear to diffusive) are considered. The cosmological
evolution of the sources, e.g.\ with parameters inspired by
observations of active galactic nuclei, also results in the same
shape of the dip. The pair-production dip is modified strongly
when the fraction of nuclei heavier than protons is high at
injection, which imposes some restrictions to the mechanisms of
acceleration operating in UHECR sources \cite{Aletal}. The shape
of acceleration spectrum needed for the $e^+e^-$-production dip
agrees with standard ones $\gamma_g=2$ for non-relativistic shock
acceleration or $\gamma_g=2.2 -2.3$ for relativistic shock. The
effective $\gamma_g= 2.6 - 2.7$ needed at ultra high energy is
naturally explained by distribution of sources over maximum energy
of acceleration or luminosity \cite{BGG,Aletal,KS}.

\begin{figure*}[ht]
\begin{center}
\includegraphics [width=0.52\textwidth]{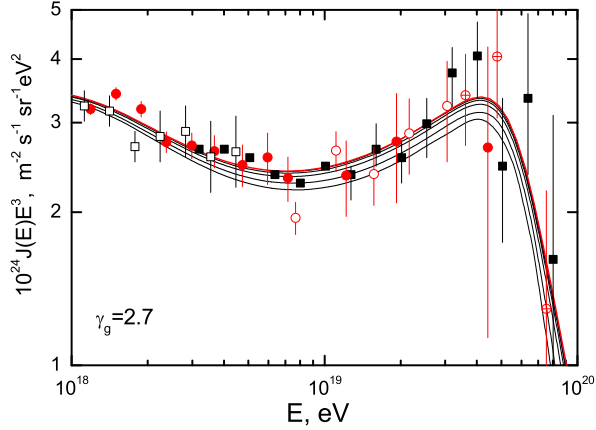}
\end{center}
\caption{The dip-based prediction for diffuse spectrum
  in energy range $ 1\times 10^{18} - 1\times 10^{20}$~eV.
  The calculated dip is normalized by calibrated
  AGASA-Yakutsk data as shown in Fig.~\ref{fig:AgHiYa}. The
  calculated diffuse
  energy spectrum is displayed for different distances $d$ between
  sources  in the range $1 - 60$~Mpc. This presents the largest
  theoretical uncertainties in energy range $(1 - 8)\times
  10^{19}$~eV. The both uncertainties in spectrum in the interval
  $(1 - 8)\times 10^{19}$~eV due to the dip-based calculations and
  measurements by AGASA, Hires and Yakutsk detectors are small enough,
  and the beginning of the GZK cutoff at $(5 - 10)\times 10^{19}$~eV
  is reliably predicted.} %
\label{fig:predict}
\end{figure*}

On the basis of the predicted dip and the calibrated data of AGASA,
HiRes and Yakutsk detectors we can plot the spectrum and flux
in the energy region $1\times 10^{18} - 1\times 10^{20}$~eV as it is
shown in  Fig.~\ref{fig:predict}.
In the energy interval $(0.1 - 10)\times 10^{19}$~eV the theoretical
uncertainties in the predicted spectrum are relatively small and
are mainly given by uncertainties in distances between sources.
These uncertainties dramatically increase at $E\gtrsim 1\times
10^{20}$ eV.

In Fig.~\ref{fig:predict} the spectra are shown for
proton-dominated flux with distances between sources in the range
$(1 - 60)$~Mpc. Therefore the beginning of the GZK cutoff
in the energy range $(5 - 10)\times 10^{19}$~eV is predicted in
the dip-based model with small uncertainties.
At larger energies the spectrum
of GZK feature is very model dependent: apart from distances
between sources it depends on fluctuations in luminosities of the
nearby sources, in  distances between them, and by maximum
acceleration energy $E_{\rm max}$ (see \cite{BGG} for calculations).
One can see from Figs.~\ref{fig:dips} and \ref{fig:predict} that the
beginning of the GZK cutoff in energy range $(5 - 10)\times
10^{19}$~eV is confirmed by all detectors, including AGASA.

\section*{Acknowledgments}
We are grateful to R. Aloisio, P. Blasi and B. Hnatyk for joint
works on this and related subjects.  We thank Alan Watson for
providing us with Auger energy spectrum  as numerical files.
Our work is supported in part by \emph{ASI} through
grant \emph{WP 1300} (theoretical study).

\end{document}